\documentclass[aps,prl,superscriptaddress,twocolumn,floatfix,showpacs]{revtex4-1}
\usepackage{graphicx,color}
\usepackage[utf8]{inputenc}
\usepackage{amsmath,amssymb,bm,amsfonts,dsfont,mathrsfs,amsthm}
\usepackage{braket}
\usepackage{txfonts,comment}
\usepackage[colorlinks=true,linkcolor=blue,citecolor=blue,urlcolor=blue]{hyperref}
\usepackage{soul}

\newcommand{\beginsupplement}
    {
    \setcounter{equation}{0}
    \setcounter{figure}{0}
    \setcounter{table}{0}
    \setcounter{page}{1}
    \makeatletter
    \renewcommand{\theequation}{S\arabic{equation}}
    \renewcommand{\thefigure}{S\arabic{figure}}
    \renewcommand{\bibnumfmt}[1]{[S##1]}
    \renewcommand{\citenumfont}[1]{S##1}
    }

\newcommand{\norm}[1]{\lVert #1 \rVert}

\begin{document}

\title{Free-Fermionic Topological Quantum Sensors}

\author{Saubhik Sarkar}
\email{saubhik.sarkar@ucalgary.ca}
\affiliation{Institute for Quantum Science and Technology and Department of Physics and Astronomy, University of Calgary, Calgary, Alberta T2N 1N4, Canada}

\author{Chiranjib Mukhopadhyay}
\email{chiranjib.mukhopadhyay@savba.sk}
\affiliation{RCQI, Institute of Physics, Slovak Academy of Sciences, Dúbravská cesta 9, 84511 Bratislava, Slovakia}
\affiliation{Institute of Fundamental and Frontier Sciences, University of Electronic Science and Technology of China, Chengdu 610051, China}

\author{Abhijeet Alase}
\email{abhijeet.alase1@ucalgary.ca}
\affiliation{Institute for Quantum Science and Technology and Department of Physics and Astronomy, University of Calgary, Calgary, Alberta T2N 1N4, Canada}

\author{Abolfazl Bayat}
\email{abolfazl.bayat@uestc.edu.cn}
\affiliation{Institute of Fundamental and Frontier Sciences, University of Electronic Science and Technology of China, Chengdu 610051, China}

\begin{abstract}

Second order quantum phase transitions, with well-known features such as long-range entanglement, symmetry breaking, and gap closing, exhibit quantum enhancement for sensing at criticality. However, it is unclear which of these features are responsible for this enhancement. To address this issue, we investigate phase transitions in free-fermionic topological systems that exhibit neither symmetry-breaking nor long-range entanglement. We analytically demonstrate that quantum enhanced sensing is possible using topological edge states near the phase boundary. Remarkably, such enhancement also endures for ground states of such models that are accessible in solid state experiments. We illustrate the results with 1D Su-Schrieffer-Heeger chain and a 2D Chern insulator which are both experimentally accessible. While neither symmetry-breaking nor long-range entanglement are essential, gap closing remains as the major candidate for the ultimate source of quantum enhanced sensing. In addition, we also provide a fixed and simple measurement strategy that achieves near-optimal precision for sensing using generic edge states irrespective of the parameter value. This paves the way for development of topological quantum sensors which are expected to also be robust against local perturbations.

\end{abstract}

\maketitle

\emph{Introduction.---}
The sensitivity of quantum systems to the variation of their environment makes them excellent sensors~\cite{degen2017quantum}. The uncertainty of measuring an unknown parameter $\lambda$, quantified by standard deviation $\delta \lambda$ is bounded by Cram\'er-Rao inequality $\delta \lambda \ge 1/\sqrt{\mathcal{M} F}$, where $\mathcal{M}$ is the number of trials and $F$ is the Fisher information~\cite{paris2009quantum}. In a classical setup, the Fisher information scales linearly with the sensor size (known as standard limit). However, quantum features, such as superposition, may enhance the resource efficiency of a quantum sensor such that the Fisher information scales quadratically with system size (known as the Heisenberg limit)~\cite{paris2009quantum}, or even faster (super-Heisenberg limit)~\cite{rams2018limits,gong2008fidelity,gu2008fidelity,greschner2013fidelity,mishra2021driving,garbe2021exponential}. There are at least two major approaches for achieving quantum enhanced sensing: (i) exploiting GHZ-type entangled states~\cite{giovannetti2004quantum, giovannetti2006quantum,frowis2011stable,demkowicz2012elusive,PhysRevA.97.042112,kwon2019nonclassicality} for estimating the angle of a unitary rotation~\cite{rams2018limits}; and (ii) utilizing quantum criticality for directly estimating the Hamiltonian parameters~\cite{zanardi2006ground, zanardi2007mixed,zanardi2008quantum,invernizzi2008optimal,gu2010fidelity,gammelmark2011phase,chu2021dynamic,liu2021experimental,montenegro2021global,skotiniotis2015quantum}.  In the former, the interaction between the particles in the quantum sensor degrades the sensing  quality~\cite{boixo2007generalized, de2013quantum,skotiniotis2015quantum,pang2014quantum}. Also, because of extreme vulnerability of GHZ states to decoherence and particle loss, it is difficult to be scaled up~\cite{kolodynski2013efficient}. In the latter, however, the interaction between the constituents of the quantum sensor is crucial and the system is more robust against decoherence. Originally the criticality-enhanced quantum sensing has been introduced for the ground state of many-body systems undergoing a second order quantum phase transition~\cite{zanardi2006ground, zanardi2007mixed, zanardi2008quantum, invernizzi2008optimal, skotiniotis2015quantum, gu2010fidelity, gammelmark2011phase, liu2021experimental, chu2021dynamic, montenegro2021global}. In such symmetry breaking transitions, the ground state reveals long-range correlations which lead to the scaling of $F\sim V^{2/D\nu}$, where $V$ is the system size (volume), $D$ is the dimension, and $\nu$ is the critical exponent with which the correlation length diverges near the criticality~\cite{rams2018limits}. Recently, quantum enhanced sensing has also been observed in integrable Floquet systems~\cite{mishra2021driving, mishra2021integrable} along the line that the Floquet gap vanishes. An important open question is what feature of a  phase transition, e.g.~symmetry breaking, long-range correlations, or vanishing gap, is truly responsible for obtaining quantum enhanced sensing.

To answer this, one needs to investigate the scaling of Fisher information in different types of quantum phase transitions, e.g., those not of the symmetry-breaking kind. Phase transitions in symmetry-protected topological (SPT)
phases of noninteracting fermions~\cite{Kitaev2009,Ryu2010} are ideal candidates for this investigation.
These topological phase transitions (TPT) are fundamentally  different from the symmetry-breaking ones in at least three aspects~\cite{Bernevig}. First, 
a fermionic SPT phase transition manifests in the form of robust edge/surface states protected against 
symmetry-preserving local perturbations~\cite{Alldridge2020,Alase}. Second, they are not detected
by a local order parameter, but rather by an integer-valued  
nonlocal quantity called a topological invariant~\cite{Bernevig}. Third, unlike the symmetry-breaking phase transitions, the 
fermionic SPT phases at TPT are short-range entangled~\cite{CGLW2013}. 
These differences between second-order quantum phase transitions and fermionic SPT phase transitions 
motivate our investigation of quantum enhanced sensitivity in the latter.
In fact, sensing based on non-Hermitian systems~\cite{Wiersig2014Enhancing} including topological systems~\cite{Schomerus2020Nonreciprocal,budich2020non, Koch2022Quantum} and TPTs for rotation angle estimation  \citep{pezze2017multipartite,zhang2018characterization,lambert2020revealing,yin2019quantum,chen2021quantum,zhang2021multipartite,yang2021superheisenberg} (like GHZ state-based metrology) have been already proposed. Nonetheless, the sensing capability of free-fermionic TPTs for estimating Hamiltonian parameters is yet to be explored. 
Fermionic SPT phases have been realized with solid-state systems~\cite{KDP1980, BHZ2006} and simulated platforms~\cite{cooper2019topological, ozawa2019topological}.
Therefore, finding quantum enhanced precision in such systems is a key step forward for developing topological quantum sensors.

In this Letter, we analytically address the quantum sensing capability of free-fermionic topological systems. We have two main findings. First, from a practical perspective, we show that these systems indeed reveal quantum enhanced sensitivity and thus are legitimate candidates for developing topological quantum sensors naturally robust against local perturbations. Second, from a fundamental perspective, we highlight the importance of gap closing, as opposed to symmetry-breaking or long-range entanglement, for quantum enhanced sensing.

\emph{Ultimate precision limit.---} 
To infer an unknown parameter $\lambda$, encoded in a quantum state $\rho_\lambda$, one has to perform measurement on the system and then feed the outcomes into an estimator algorithm. For a basic introduction to single parameter estimation, we refer to  Supplemental Material (SM)~\cite{Supp}. For a given measurement setup, described by a set of projective operators $\{ \Pi_n\}$, every outcome appears with the probability $p_n(\lambda)=\text{Tr}\left[\rho_\lambda \Pi_n\right]$. In this case, all the information is encoded in a classical probability distribution and thus the Cram\'er-Rao bound is determined by classical Fisher information (CFI), defined as $F^C{=}\sum_k p_n (\partial_\lambda\log p_n)^2$. One can maximize the CFI for all possible measurement setups to obtain quantum Fisher information (QFI) as the ultimate precision bound. The QFI can be computed as $F^Q=\text{Tr}\left[\mathcal{L}_{\lambda}^2 \rho_\lambda \right]$, where $\mathcal{L}_\lambda$ is the symmetric logarithmic derivative (SLD) operator defined as $\partial_{\lambda}\rho_{\lambda}=(\rho_\lambda \mathcal{L}_\lambda+\mathcal{L}_\lambda \rho_\lambda)/2$. For pure states $\rho_{\lambda} = \ket{\psi_{\lambda}} \bra{\psi_{\lambda}}$, SLD operator simplifies to $\mathcal{L}_{\lambda} = 2 \partial_{\lambda} \rho_{\lambda}$, and $F^Q = 4\left(\braket{\partial_\lambda \psi_{\lambda}|\partial_\lambda \psi_{\lambda}} - |\braket{\partial_\lambda \psi_{\lambda}|\psi_{\lambda}}|^2 \right)$ \citep{paris2009quantum}. It is worth emphasizing that the optimal  measurement setup that achieves the ultimate precision bound is not unique, although one solution is always given by the eigenvectors of the SLD operator. 

\emph{Free-fermionic SPT model.---}
Free-fermionic SPT phases host energy excitations localized on the boundary known as edge/surface states. 
The existence of these states is guaranteed by the nontrivial 
topology of the filled band wave functions~\cite{Alldridge2020}.
These states are studied using tight-binding models~\cite{acov16,acov17,caov17,caov18} (see SM~\cite{Supp} for more details).
We first analyze the QFI of the edge states in 1D systems and later generalize our results to higher dimensions.
Consider a 1D lattice with sites labeled by $\{j: j {\in} [L]\}$, where $[L] {=} \{0,1,\dots,L-1\}$. 
Suppose there are $d$ internal degrees of freedom associated to each lattice site.
The single-particle Hilbert space $\mathscr{H}$ is then spanned by orthonormal basis states $\{\ket{j,m}: j {\in} [L], m {\in} [d]\}$,
and can be tensor-factorized as
$\mathscr{H} {\cong} \mathscr{H}_L {\otimes} \mathscr{H}_I$~\cite{acov16,acov17}.
The single-particle Hamiltonian of a number-conserving, noninteracting fermionic system with uniform 
coupling and open boundary conditions (OBC) can be expressed as
\begin{equation}
\label{HamOBC}
    H {=} \sum_{j \in [L]}\ket{j}\bra{j}\otimes h_{0}(\lambda) + 
    \sum_{j < j' \in [L]}\left(\ket{j}\bra{j'}\otimes h_{j'-j}(\lambda) + \text{H.c.}\right) \,,
\end{equation}
where each $h_{j'-j}$ is a $d {\times} d$ matrix whose entries are complex amplitudes of 
hopping between lattice sites separated by distance $j'{-}j$ possibly accompanied by a change in 
internal state~\cite{acov16,acov17}, and these entries
depend on the parameter $\lambda$ that is being estimated. 
Hamiltonians of the form in Eq.\,\eqref{HamOBC} are routinely used for
investigation of edge states in 1D free-fermionic topological systems. 
Our analysis can be generalized to number nonconserving systems including Kitaev chain~\cite{kitaev2001unpaired}, 
as Bogoliubov-de Gennes Hamiltonian of such systems has the same structure as in Eq.~\eqref{HamOBC}.

\emph{Edge states in 1D.---}
The zero-energy edge states (topologically protected or accidental) 
localized on the $j=0$ edge are described (or well-approximated) by 
$\ket{\psi_{\rm edge}} = \ket{\phi_z}\ket{u}$, where
\begin{equation}
\label{generic_edge}
    \ket{\phi_z}  = \sqrt{\frac{1-|z|^2}{1-|z|^{2L}}} \sum_{j \in [L]}z^j\ket{j}, \quad z\in \mathbb{C},\ |z|<1 
\end{equation}
parametrized by $z$ accounts for the spatial part of exponentially decaying nature of $\ket{\psi_{\rm edge}}$, and 
$\ket{u} \in \mathscr{H}_I$ is an internal state vector~\cite{acov16,acov17,caov17}.
Both $z$ and $\ket{u}$ depend on $\lambda$ in general. 

We now derive the scaling of QFI for $\ket{\psi_{\rm edge}}$, assuming $\arg(z)$ is 
independent of $\lambda$,
and leave the general case for SM~\cite{Supp}.
Our results are stated using $O$, $\Omega$, and $\Theta$ asymptotic 
notations, to denote upper, lower, and tight bounds on the scaling respectively~\cite{brassard}. 
The QFI for $\ket{\psi_{\rm edge}}$ with respect to $\lambda$ 
can be expressed as
$F_{\ket{\psi_{\rm edge}}}(\lambda) {=} F_{\ket{\phi_z}}(\lambda) + F_{\ket{u}}(\lambda)$. 
For $L {\gg} 1$, both $z$ and $\ket{u}$ approach a fixed value that does not depend on $L$~\cite{acov16,acov17}.
However, the state $\ket{\phi_z}$ depends on $L$ due to the
normalization. Therefore, the scaling of QFI comes from
the scaling of $F_{\ket{\phi_z}}(\lambda) {=} 4\left(\braket{\partial_\lambda \phi_z|\partial_\lambda \phi_z} - |\braket{\partial_\lambda \phi_z|\phi_z}|^2 \right)$.
For $\arg(z)$ independent of $\lambda$,  $\braket{\partial_\lambda \phi_z | \phi_z}{=}0$, and simple algebra reveals
\begin{equation}
    F_{\ket{\phi_z}}(\lambda) = \frac{4(\partial_{\lambda} |z|)^2[1+|z|^{4L}-|z|^{2L-2}(2|z|^2+L^2(1-|z|^2)^2)]}{(1-|z|^2)^2(1-|z|^{2L})^2} \, .
    \label{Fz}
\end{equation}
Away from TPT, $|z|{<}1$ yields $\text{lim}_{L \to \infty} F_{\ket{\phi_z}}(\lambda) = 4(\partial_{\lambda} z)^2(1-z^2)^{-2}$, so that
$F_{\ket{\psi_{\rm edge}}}(\lambda) \in \Theta(1)$. 
As edge states are localized single particle excitations, we do not expect $L$-dependent scaling away from TPT. In contrast, at TPT,
the zero-energy edge states undergo delocalization, so that $|z| \to 1$
as $\lambda$ approaches the transition point $\lambda_{\rm c}$~\cite{acov16,acov17}. Consequently,
by calculating the limit of Eq.\,\eqref{Fz} as $\lambda \to \lambda_{\rm c}$, we get
\begin{equation}
    \lim_{\lambda \to \lambda_{\rm c}} F_{\ket{\phi_z}}(\lambda) {=} \frac{(\partial_{\lambda} z)^2(L^2 {-} 1)}{3}
    {\implies}  F_{\ket{\psi_{\rm edge}}}(\lambda_{\rm c})\in \Theta(L^2) \,,
\end{equation}
independent of the model Hamiltonian. The same result holds for complex $z$, as we show in the SM~\cite{Supp}.
This quadratic scaling of the QFI of edge states at the phase transition is a remarkable observation showing the power of free-fermionic topological systems for achieving quantum enhanced sensitivity. This is in fact the first main result of our work.

\emph{Edge states in higher dimensions.---}
We now investigate the scaling of the QFI of the edge states of $D$-dimensional systems in which
periodic boundary conditions (PBC) are enforced along $D {-} 1$ directions, and OBC along the remaining 
direction. For ease of explanation, consider a 2D square lattice with the orthonormal basis states
$\{\ket{j_1,j_2,m}: j_1 \in [L_1], j_2 \in [L_2], m \in [d]\}$~\cite{caov18}.
A Hamiltonian with PBC along both the spatial directions can be expressed as
$H_{\rm PBC} {=} \oplus_{\boldsymbol{k}}H_{\boldsymbol{k}}$ with $\boldsymbol{k}$ in the 2D Brillouin zone and $H_{\boldsymbol{k}}$ the Bloch Hamiltonian. 
If OBC is enforced along the first spatial direction, then $\boldsymbol{k}$ is no longer a good quantum number. However, $k_\parallel$ (component of $\boldsymbol{k}$ along the periodic direction) remains a good quantum number, and therefore the total Hamiltonian can be expressed as $H_{\rm OBC} {=} \oplus_{k_\parallel}H_{k_\parallel}$, with $H_{k_\parallel}$
denoting the Hamiltonian of a virtual 1D wire labeled by $k_\parallel$~\cite{caov18}.
Each $H_{k_\parallel}$ has a structure similar to that of the Hamiltonian in Eq.\,\eqref{HamOBC}.
An edge state $\ket{\psi_{\rm edge}}$ at a fixed $k_\parallel$ is well-approximated by 
$\ket{\psi_{\rm edge}} {=} \ket{k_\parallel}\ket{\phi_z}\ket{u(z,k_\parallel)}$,
where $\ket{\phi_z}$ is given in Eq.~\eqref{generic_edge} with the replacement $L {\to} L_2$, and $\ket{k_\parallel} {=} \frac{1}{\sqrt{L_1}}\sum_{j_1 \in [L_1]}e^{ik_\parallel j_1}\ket{j_1}$ with $k_\parallel {\in} [-\pi,\pi)$. 
For an edge state at a fixed value of $k_\parallel$, we have 
$\partial_{\lambda}\ket{k_\parallel} {=} 0$, and therefore
$F_{\ket{\psi_{\rm edge}}}(\lambda) = F_{\ket{\phi_z}}(\lambda) + \text{constant} \in \Theta(L_2^2)$ at TPT as in the 1D case. 
Interestingly, for $L_1 {=} L_2 {=} L$, we have $F_{\ket{\psi_{\rm edge}}}(\lambda) \in \Theta(V)$ where $V {=} L^2$ is the total system size (area).
In $D$-dimensions, lattice sites are indexed by $\{j_1,\dots,j_D\}$, and PBC are
enforced along the first $D {-} 1$ directions. Similar analysis as above yields
$F_{\ket{\psi_{\rm edge}}}(\lambda) \in \Theta(V^{2/D})$ at TPT, similar to the behaviour of QFI at second order phase transitions~\cite{rams2018limits}. This establishes the scaling of QFI of edge states in any spatial dimension.

\emph{Optimal measurement basis for edge states.---} 
While QFI determines the ultimate precision bound, its saturation in the Cram\'er-Rao inequality relies on the choice of an optimal measurement basis. For the case where $\arg(z)$ is independent of $\lambda$, the position measurement in the basis
\begin{equation}
    \mathcal{B} = \left \lbrace \ket{j}\bra{j}\otimes \mathds{1}_d,\   
    j\in [L]  \right \rbrace
    \label{basis}
\end{equation}
is sufficient to saturate the Cram\'er-Rao bound for QFI $F_{\ket{\phi_z}}(\lambda)$ for every $\lambda$. We note that for the generic QFI expression $F_{\ket{\psi_{\rm edge}}}(\lambda) {=} F_{\ket{\phi_z}}(\lambda) + F_{\ket{u}}(\lambda)$, only the first term contributes to scaling. The second term $F_{\ket{u}}$, coming from the intrasite physics, does not depend on lattice size and approaches a fixed value for $L \gg 1$. Consequently, the measurement of $\ket{\psi_{\rm edge}}$ performed in this basis yields optimal precision up to a length-independent constant. The proof, obtained by showing that QFI for $\ket{\phi_z}$ equals the CFI in this basis, is detailed in the SM \citep{Supp}. Physically, this entails measuring the location of the particle in the lattice. We emphasize that this measurement basis, independent of parameter values and obviously site-local, is in sharp contrast to many proposals of quantum many-body sensors, where Heisenberg scaling is achievable in theory, but only through highly non-local, complicated, and parameter-sensitive measurement bases. Even dropping the assumption of $\arg(z)$ being independent of $\lambda$, we show in the SM \citep{Supp} that this measurement basis still yields quadratic scaling of QFI, which lends generality to quantum-enhanced sensitivity.

At this point we note that the localization feature of the edge states is not necessary for quadratic scaling of QFI at TPT. For example, as we later show numerically, quadratic scaling is observed for the bulk states at the top of the lower energy band as well as bottom of the upper energy band as we approach TPT from the trivial phase. However, for such states, the  measurement basis in Eq.\,\eqref{basis} may not be optimal.

\begin{figure}[t]
\centering
  \begin{tabular}{cc}
    \includegraphics[width=0.45\linewidth]{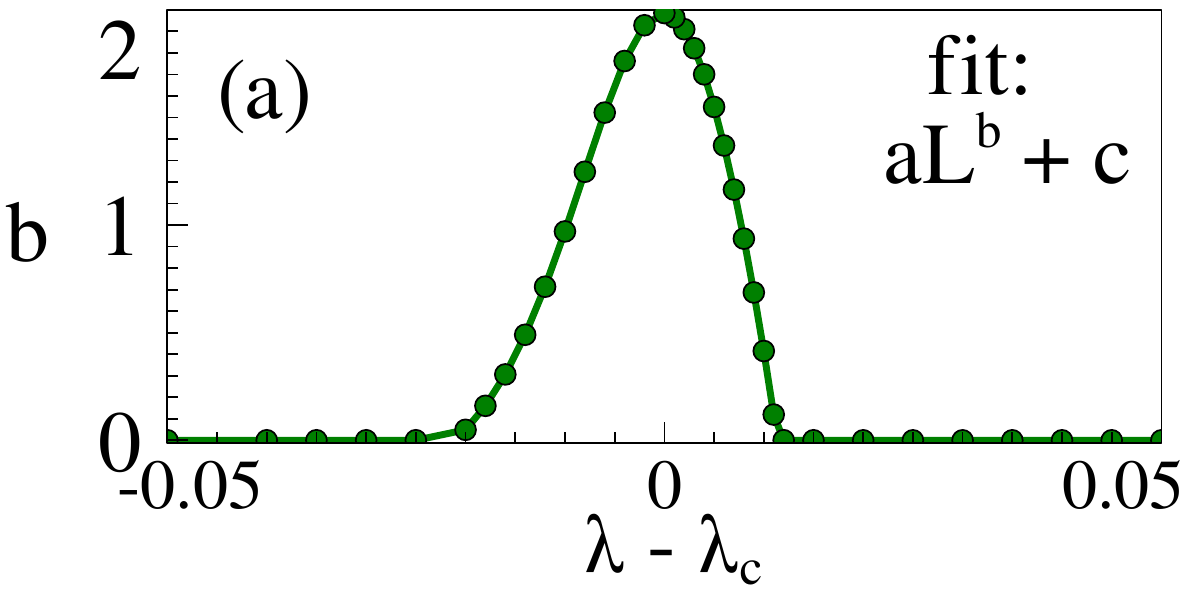} 
    \includegraphics[width=0.45\linewidth]{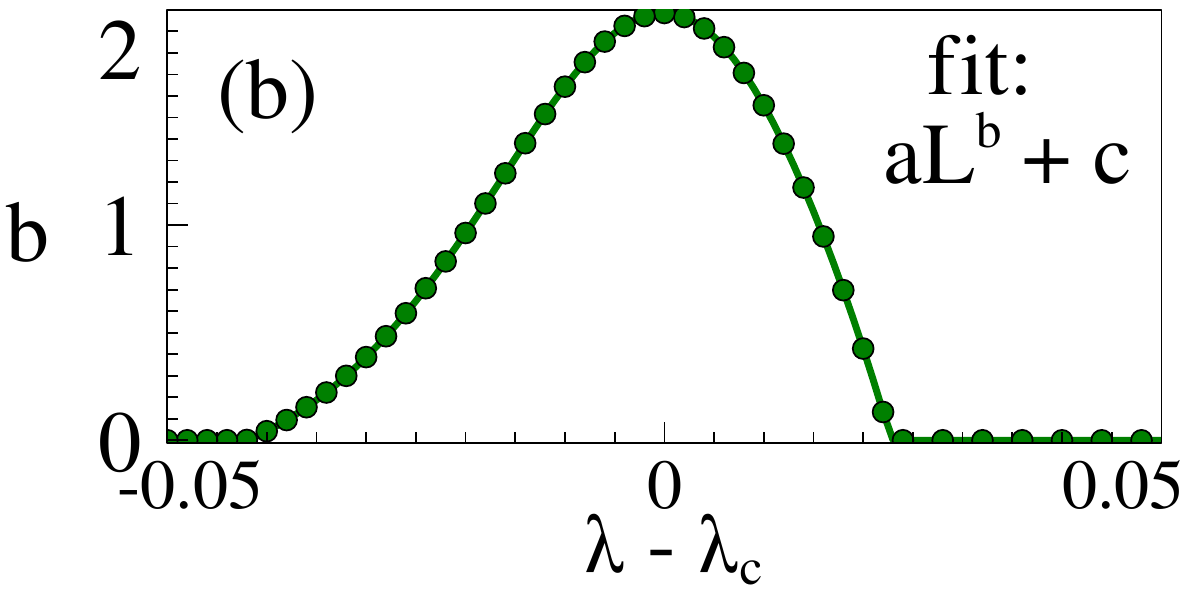}
  \end{tabular}
\caption{Scaling exponent of QFI of edge state (for $\lambda < \lambda_{\rm c}$) and corresponding bulk state (for $\lambda > \lambda_{\rm c}$) as a function of $\lambda$ for (a) SSH model, and (b) Chern insulator model.}
\label{fig_edge}
\end{figure}

\emph{Example for 1D.---}
As a concrete example we consider the Su-Schrieffer-Heeger (SSH) Hamiltonian~\cite{PhysRevLett.42.1698} 
\begin{align}
    \hat{H}^{\rm SSH} = -\sum_{j \in [L]} \left( J_1 \hat{b}_j^{\dag} \hat{a}_j+ J_2 \hat{a}_{j+1}^{\dag} \hat{b}_j + \text{H.c.}\right) \,,
    \label{SSHham}
\end{align}
where $J_1$ and $J_2$ are the exchange couplings of the two internal states (denoted by fermionic operators $\hat{a}_j$ and $\hat{b}_j$ at site $j$) at a single site and between adjacent sites, respectively. This Hamiltonian is of the same form as in Eq.~\eqref{HamOBC}, with only nonzero matrices $h_0 {=} -J_{1} \sigma_x$, and $h_1 {=} -J_2 ( \sigma_{x} -i \sigma_{y})/2$, with $\sigma_x, \sigma_y$ being Pauli matrices. This model exhibits TPT at $J_1 {=} J_2$ protected by sublattice symmetry, and
has been realized in both solid state~\cite{PhysRevLett.42.1698} and optical lattice~\cite{atala2013direct} experiments. 
For simplicity, we shall assume that $\hat{b}_{L-1}$ is isolated from other sites in the SSH chain ~\footnote{This assumption leads to special boundary conditions as discussed in Appendix E of Ref.~\cite{caov18}. These boundary conditions are different from the open boundary conditions discussed in Eq.\,\eqref{HamOBC}, but it does not change the form of edge states in Eq.\,\eqref{generic_edge}.}, and $\lambda {=} J_1/J_2$ is a real parameter which has to be estimated. For $\lambda {<} \lambda_{\rm c}=1$, the normalized edge state solution is given by~\cite{zaimi2021detecting} $ |\psi_{\rm edge} ^{\rm SSH}\rangle {=} \ket{\phi_{z = -\lambda}}\ket{u}$, where $\ket{\phi_z}$ is given by Eq.~\eqref{generic_edge} and $\ket{u} = [1 \enspace 0]^T$. We obtain the same scaling relation for QFI as in the general case. To verify this numerically, one can use a fit function $aL^b+c$ to the QFI of $\ket{\psi_{\rm edge} ^{\rm SSH}}$ for each value of $\lambda$ and extract the exponent $b$. 
For the trivial regime ($\lambda > \lambda_{\rm c}$), the edge state smoothly deforms into the bulk state at the top of the lower energy band, which we use to calculate the QFI scaling. As displayed in Fig.~\ref{fig_edge}(a), scaling of QFI changes from quadratic ($b{=}2$) to constant ($b{=}0$), as one moves away from TPT. Note that $z$ is real and $\ket{u}$ is constant, therefore the measurement described by Eq.~\eqref{basis} is optimal.


\emph{Example for 2D.---}
We now illustrate the scaling of QFI for a Chern insulator on a square lattice -- a prototype of topological insulators with broken time-reversal symmetry~\cite{Ryu2010}. It has also been experimentally realized in optical lattices~\cite{wu2016realization}. The spin-orbit coupled (SOC) Hamiltonian is~\cite{sticlet2012geometrical, zhang2017detecting} 
\begin{align}
\label{Chham}
    \hat{H}^{\rm Ch} = \displaystyle\sum_{\boldsymbol{k}} \begin{bmatrix} \hat{c}^{\dag}_{\boldsymbol{k},\uparrow} & \hat{c}^{\dag}_{\boldsymbol{k},\downarrow}\end{bmatrix} \enspace H_{\boldsymbol{k}}^{\rm Ch} \enspace
    \begin{bmatrix} \hat{c}_{\boldsymbol{k},\uparrow} & \hat{c}_{\boldsymbol{k},\downarrow}\end{bmatrix} ^T \,,
\end{align}
where $H_{\boldsymbol{k}}^{\rm Ch} = \boldsymbol{B} \cdot \boldsymbol{\sigma}$ is the Bloch Hamlitonian with $\boldsymbol{B} = \left(2 t_1 \cos{k_x}, \, 2 t_1 \cos{k_y}, \, m_z + 2 t_2 (\sin{k_x} + \sin{k_y})\right)$ and $\boldsymbol{\sigma}$ the vector of Pauli matrices. 
Here $\uparrow,\downarrow$ denote spin-$1/2$ up and down states, and $m_z, t_1, t_2$ are lattice parameters. We will consider $\lambda{=}m_z/t_2$ as the parameter to be estimated. The eigenvectors form two bands that touch at phase transition at the Dirac points $(k_x, k_y) {=} \pm(\pi/2, \pi/2)$ for nonzero $\lambda$, and the phase boundaries are given by $\lambda_{\rm c} = \mp 4$~\cite{zhang2017detecting}. 
We impose PBC along $x$ direction ($k_\parallel = k_x$), and 
decompose the Hamiltonian as $H^{\rm Ch} {=} \oplus_{k_x}H_{k_x}$,
where $H_{k_x}$ describes a virtual 1D wire Hamiltonian of the form in Eq.\,\eqref{HamOBC} with $ h_0(k_x) = 2 t_1 \cos{k_x}\sigma_x + (m_z + 2 t_2 \sin{k_x})\sigma_z, h_1(k_x) = t_1\sigma_y - it_2\sigma_z $.
The QFI of the edge state at $k_x = \pi/2$ localized near $j=0$, displayed in Fig.~\ref{fig_edge}(b), shows quadratic scaling at TPT and constant scaling away from it. As before, the quadratic scaling is shown to be approached from the trivial phase as well for the corresponding bulk state.

\emph{QFI of many-body ground state.---}
We now look at the scaling nature for the fermionic many-body ground states, which are relevant for solid state experiments.
We first derive a formula for the QFI of a general many-body state $\ket{\Psi}$ of $N$ fermions occupying single-particle states denoted by $\ket{\psi_1}, \dots, \ket{\psi_N}$.
The antisymmetrized wave function for this state is given by the Slater determinant formula~\cite{fetter} $\ket{\Psi} = (1/\sqrt{N!})\sum_{\sigma \in {S_N}}\text{sgn}(\sigma)\ket{\psi_{\sigma_1}}\dots\ket{\psi_{\sigma_N}}$ where $S_N$ is the symmetric group.
The QFI of this state, with $P = \sum_{l=1}^{N}\ket{\psi_{l}}\bra{\psi_{l}}$ as projector on the occupied states, simplifies to (see SM~\cite{Supp})

\begin{align}
\label{mbqfi}
    F_{\ket{\Psi}} =4\sum_{l=1}^{N}\braket{\partial_\lambda \psi_{l}|\mathds{1} - P|\partial_\lambda \psi_l} .
\end{align}

We now analytically derive the scaling of QFI under PBC, and later numerically validate similar results for OBC. Consider the ground state in the $D$-dimensional case under PBC, with filled lowest band and empty higher bands. Translational invariance dictates that each single-particle state in the filled band is of the form $\ket{\psi_{\boldsymbol{k}}} {=} \ket{\boldsymbol{k}} \ket{u_{\boldsymbol{k}}}$, where $\ket{\boldsymbol{k}}$ is the plane-wave state. Using $\ket{\partial_\lambda \psi_{\boldsymbol{k}}} {=} \ket{\boldsymbol{k}} \ket{\partial_\lambda u_{\boldsymbol{k}}}$, Eq.~\eqref{mbqfi} simplifies to
\begin{align}
\label{manybodyqfiperiodic}
    F_{\rm GS}^{\rm PBC} {=} 4 \sum_{\boldsymbol{k}} \left(\braket{\partial_\lambda u_{\boldsymbol{k}} | \partial_\lambda u_{\boldsymbol{k}}} - |\braket{\partial_\lambda u_{\boldsymbol{k}} | u_{\boldsymbol{k}}}|^2 \right) {=} \sum_{\boldsymbol{k}} F_{\ket{u_{\boldsymbol{k}}}} \,.
\end{align}

\emph{Many-body QFI at TPT.---}
We now show how $\Omega(L^2)$ scaling of $F_{\rm GS}^{\rm PBC}$ emerges
at TPT in a simplistic model of band-gap inversion in 1D systems~\cite{Bernevig}. 
Consider a two-band Hamiltonian that can be approximated as 
$ H_k = \alpha k \sigma_x + (\lambda-\lambda_{\rm c})\sigma_z \ $ near the Dirac point $k=0$,
with $\alpha$ a Hamiltonian parameter independent of $\lambda$ and $L$.
The two energy bands touch at $k{=}0$ at TPT (i.e.~$\lambda {=} \lambda_{\rm c}$).
Major contribution to QFI is expected from the states near the Dirac point.
In fact, as shown in SM~\cite{Supp}, it is enough to consider only the two lowest $k$-states to establish a $\Theta(L^2)$ scaling for QFI. This, combined with Eq.\,\eqref{manybodyqfiperiodic}, rules out subquadratic scaling of $F_{\rm GS}^{\rm PBC}$, so that $F_{\rm GS}^{\rm PBC} \in \Omega(L^2)$.

To explicitly see this scaling behaviour at TPT we look at our prototypical examples of 1D SSH chain and 2D Chern insulator mentioned before. In the first case (see SM~\cite{Supp})
\begin{align}
     F_{\rm PBC}^{\rm SSH} (\lambda_{\rm c}) =  \sum_{\kappa=1}^{L-1} \frac{\cot^2(\pi \kappa/L)}{4} = \frac{L^2-3L+2}{12} \,,
\end{align}
which clearly shows the $\Theta(L^2)$ scaling for large $L$.
Moreover, the Fock basis is an optimal measurement basis, 
as the ground state of SSH Hamiltonian has real coefficients in that basis~\cite{zaimi2021detecting}.
Such a measurement can be performed by measuring the number operator $\hat{c}^\dagger_{j,m} \hat{c}_{j,m}$ for each fermionic mode.

For the Chern insulator on a $L {\times} L$ lattice, QFI at TPT is~\cite{Supp}
\begin{align}
    F_{\rm PBC}^{\text{Ch}} (\lambda_{\rm c}) = \displaystyle\sum_{\boldsymbol{k} \ne (\pi/2, \pi/2)} \frac{B_x^2 + B_y^2}{4 (B_x^2 + B_y^2 + B_z^2)^2} \,.
\end{align}
As we show later, this sum also shows $\Omega(L^{2})$ dependence.

\emph{Many-body QFI away from TPT.---}
To see the scaling of $F_{\rm GS}^{\rm PBC}$ away from TPT, we prove that $F_{\ket{u_k}}$ is bounded by a constant independent of $N$.
Therefore, $F_{\rm GS}^{\rm PBC} \in O(N)$, by Eq.~\eqref{manybodyqfiperiodic}. 
We first observe that 
$F_{\ket{u_k}} \le 4\braket{\partial_\lambda u_k|\partial_\lambda u_k}$. First-order perturbation theory yields $ \braket{\partial_\lambda u_k|\partial_\lambda u_k} = |\braket{v_k|\partial_\lambda H|u_k}|^2 / (\epsilon_{1,k} - \epsilon_{0,k})^2$,
where $\ket{v_k}$ is the higher band wavefunction, and $\epsilon_{0,k}$ ($\epsilon_{1,k}$) is the 
lower (higher) band energy eigenvalues. Now we can bound $F_{\ket{u_k}}$ by $4\braket{\partial_\lambda u_k|\partial_\lambda u_k} \le 4\norm{\partial_\lambda H_k}^2 / \Delta E^2$, where $\Delta E$ is the band gap, and $\norm{\bullet}$ denotes the operator norm. Furthermore, $\norm{H_k} \le \text{sup}\norm{\partial_\lambda H_k} = \norm{\partial_\lambda H}$, hence $\braket{\partial_\lambda u_k|\partial_\lambda u_k} \le \norm{\partial_\lambda H}^2 / \Delta E^2$. This proves that QFI of $\ket{\Psi^{\rm PBC}_{\rm GS}}$ scales at most linearly with the system size ($O(N) = O(V)$) away from TPT. This is in stark contrast with the constant scaling of QFI for the edge states.  

\begin{figure}[t]
\centering
  \begin{tabular}{cc}
    \includegraphics[width=0.48\linewidth]{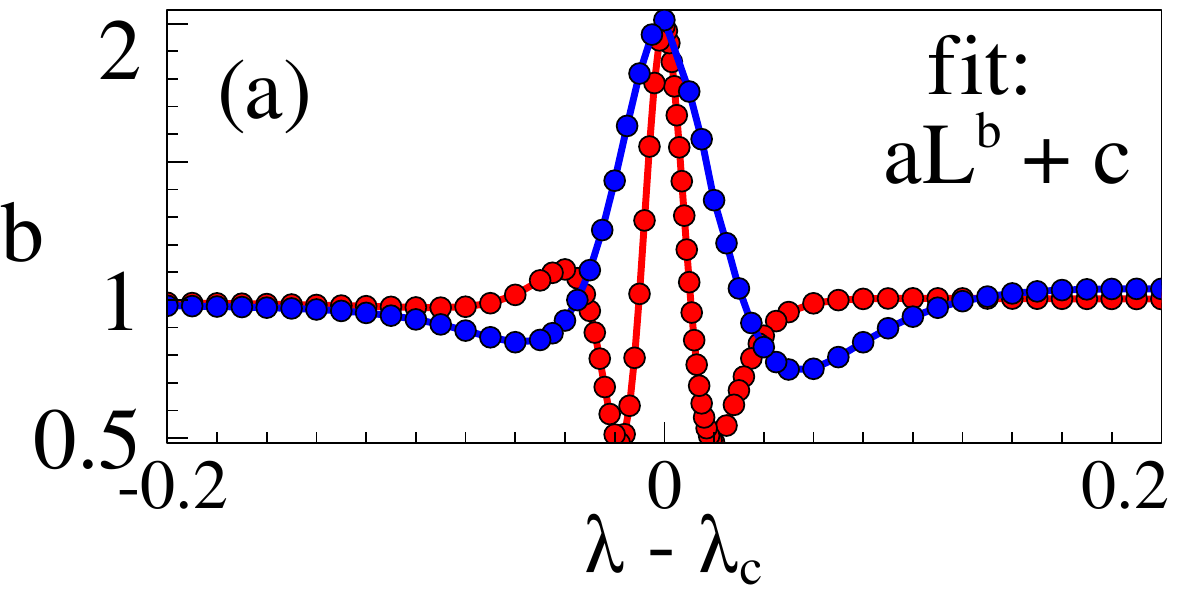} 
    \includegraphics[width=0.48\linewidth]{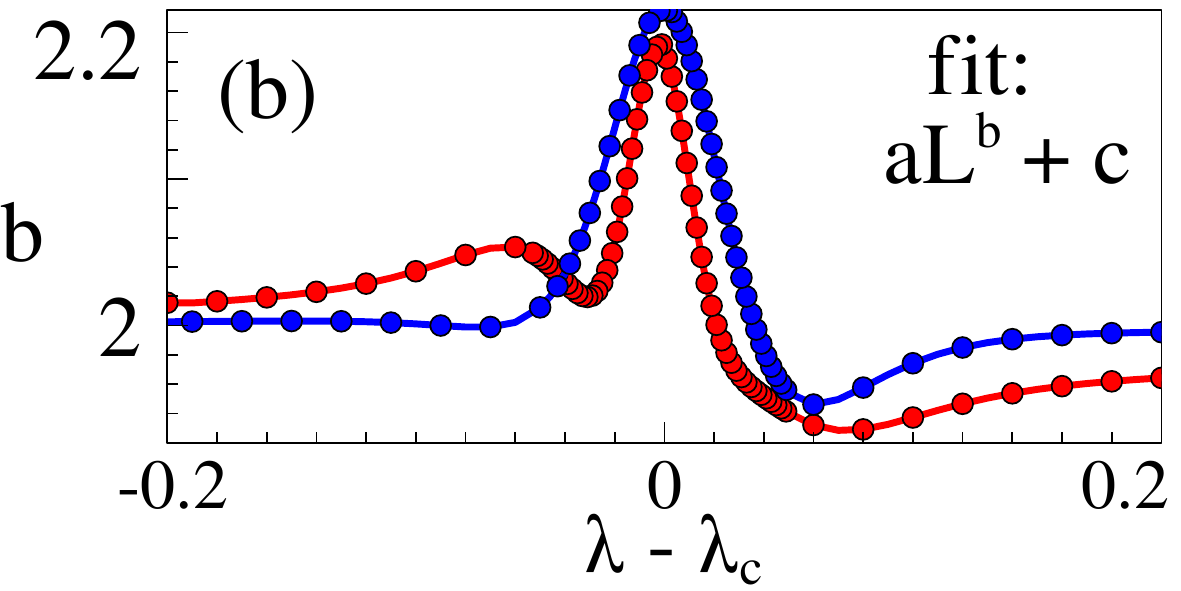}
  \end{tabular}
\caption{Scaling exponent of QFI of many-body ground state as a function of $\lambda$ subject to PBC (blue) and OBC (red) for (a) SSH model, and (b) Chern insulator model.}
\label{fig_exponent}
\end{figure}

Linear scaling away from TPT can be explicitly proved for SSH model in the continuum limit $L {\rightarrow} \infty$  (see SM~\citep{Supp}), as 
\begin{equation}
    \lim_{L\to \infty} \frac{F_{\rm PBC}^{\text{SSH}}(\lambda)}{L} = \begin{cases}
       1/2(1-\lambda^2) &\quad\text{if } \lambda < 1\\ 
       1/2(\lambda^4 - \lambda^2) &\quad\text{if } \lambda > 1 .\\ 
     \end{cases}
\end{equation}
We further provide numerical confirmations by repeating the fitting procedure as before. The scaling exponents versus $\lambda$ are shown in Figs.~\ref{fig_exponent}(a)-(b) for the SSH and Chern insulator model respectively. Expectedly, the $\Omega(L^2)$ scaling at TPT and $O(L^D)$ scaling far enough away in the topological phase are independent of the boundary conditions. We observe qualitatively similar scaling behaviour in the trivial phase as well. For the Chern insulator, true OBC are numerically intractable beyond small system sizes, hence we use strip geometry, which leads to the small discrepancies with the PBC results. 

\emph{Experimental realization.---}
All the ingredients for our proposals are already present in cold atom experiments in optical lattices. As edge states are single-particle states they have been observed with both fermions~\cite{Mancini2015Observation} and bosons~\cite{Stuhl2015Visualizing} in quantum Hall systems on optical lattices using standard imaging techniques for synthetic dimensions by populating the edge states without populating the bulk. For SSH chain, proposals for edge state preparation are also in place~\cite{Bar13, Krivosenko2018Resonant}. Position basis for optimal measurement can be accessed using quantum gas microscopy~\cite{Simon2011Quantum}. To access the fermionic many-body state filling up the entire lower band one can bank on the successful experiments on 1D SOC lattice systems~\cite{Cheuk2012Spin, Livi2016Synthetic, Kolkowitz2017Spin}. For 2D cases, the fermionic lattice Hamiltonians are yet to be realized but SOC has been observed in trapped gases~\cite{Huang2016Experimental, Meng2016Experimental}. 

\emph{Conclusion.---}
Through analytical investigation, we show that one can achieve precision beyond the standard limit at the transition point of  free-fermionic topological models. This paves the way for development of topological quantum sensors, which are  expected to be robust against local perturbations. 
Our edge-state based schemes allow achieving Heisenberg-limited sensing via a simple position measurement, thus avoiding the necessity of complicated highly entangled optimal measurements that hitherto seemed necessary to build quantum many-body sensors. From a fundamental point of view, our analysis indicates that gap closing, rather than long-range entanglement and spontaneous symmetry-breaking, is essential for obtaining quantum enhanced precision. This observation is consistent with recent discovery of quantum enhanced sensitivity at Floquet gap closing~\cite{mishra2021driving, mishra2021integrable} in periodically driven systems.

\begin{acknowledgments}
A.~B.~acknowledges support from the National Key R\&D Program of China (Grant No.~2018YFA0306703), National Science Foundation of China (Grants No.~12050410253 and No.~92065115) and the Ministry of Science and Technology of China (Grant No.~QNJ2021167001L). S.~S.~acknowledges support by Alberta Major Innovation Fund. C.~M.~acknowledges Slovak Academy of Sciences for funding from OPTIQUTE APVV-18-0518 and DESCOM VEGA-2/0183/21 projects, and the Stefan Schwarz Support Fund. A.~A.~acknowledges support by Killam Trusts (Postdoctoral Fellowship).
\end{acknowledgments}

\bibliographystyle{apsrev4-1}
\bibliography{TQS}

\clearpage
\onecolumngrid
\beginsupplement
\section{Supplemental Material: F\texorpdfstring{\MakeLowercase{ree-}}{}F\texorpdfstring{\MakeLowercase{ermionic}}{} T\texorpdfstring{\MakeLowercase{opological}}{} Q\texorpdfstring{\MakeLowercase{uantum}}{} S\texorpdfstring{\MakeLowercase{ensors}}{}}

\begin{center}
\vspace*{0.2cm}
Saubhik Sarkar,$^{1}$ Chiranjib Mukhopadhyay,$^{2,3}$ Abhijeet Alase,$^{1}$ and Abolfazl Bayat$^{3}$ \\
\vspace*{0.2cm}
$^{1}${\small \em Institute for Quantum Science and Technology and Department of Physics and Astronomy,} \\
      {\small \em University of Calgary, Calgary, Alberta T2N 1N4, Canada} \\
$^{2}${\small \em RCQI, Institute of Physics, Slovak Academy of Sciences, Dúbravská cesta 9, 84511 Bratislava, Slovakia} \\
$^{3}${\small \em Institute of Fundamental and Frontier Sciences, University of Electronic Science and Technology of China, Chengdu 610051, China} \\
\end{center}

\subsection{Quantum single parameter estimation}
Here we give a brief overview of sensing of a single parameter $\lambda$ encoded in a quantum system at state $\rho_\lambda$ in terms of the QFI of the system. To do so, let us first consider a statistical system with (classical) probability distribution $p(x|\lambda)$ over some random variable $x$ with outcomes $\lbrace x_i\rbrace$, when the actual parameter value is $\lambda$. If one chooses an estimator $\hat{\xi} (x)$ for the parameter $\lambda$ such that the estimator is unbiased, then this means
\begin{align}
& E[\hat{\xi} (x)]  = \lambda \nonumber \\
& \Rightarrow\sum_{i} \left( \hat{\xi}(x_i) - \lambda \right) p(x_i|\lambda) = 0 \nonumber \\
& \Rightarrow \partial_{\lambda} \left[\sum_{i} \left( \hat{\xi}(x_i) - \lambda \right) p(x_i|\lambda)\right] = 0 \nonumber \\
& \Rightarrow \sum_{i}  \left( \hat{\xi}(x_i) - \lambda \right) \frac{\partial p(x_i|\lambda)}{\partial \lambda} - \sum_i p(x_i|\lambda) = 0 \nonumber \\
& \Rightarrow \sum_{i}  \left( \hat{\xi}(x_i) - \lambda \right) \frac{\partial p(x_i|\lambda)}{\partial \lambda} = 1 \nonumber \\
& \Rightarrow \sum_{i}  \left( \hat{\xi}(x_i) - \lambda \right)  p(x_i|\lambda) \frac{\partial \log(p(x_i|\lambda)}{\partial \lambda} = 1 
\end{align}
Now, we make use of the Cauchy-Schwarz inequality $\sum_{i} a_i^2 \sum_{i} b_i^2 \geq (\sum_i a_i b_i)^2$, where we put $a_i = \left( \hat{\xi}(x_i) - \lambda \right) \sqrt{p(x_i|\lambda)}$ and $b_i = \sqrt{p(x_i|\lambda)} \partial \log(p(x_i|\lambda)/\partial \lambda $. From this, we immediately obtain the following lower bound  for $\text{Var}[\hat{\xi}(x)] = \sum_i \left( \hat{\xi}(x_i) - \lambda \right)^2 p(x_i|\lambda) $
\begin{equation}
\text{Var}[\hat{\xi}(x)] \geq \frac{1}{\sum_{i} p(x_i|\lambda) \left( \log p(x_i|\lambda) \right)^2} = \frac{1}{F^C(\lambda)},
\end{equation}
where $F^C$ is the (classical) Fisher information of the system. This inequality is known in statistics literature as the Cramer-Rao inequality. For single parameters, it is easy to note that this inequality is saturated whenever $a_i = \left( \hat{\xi}(x_i) - \lambda \right) \sqrt{p(x_i|\lambda)}$ and $b_i = \sqrt{p(x_i|\lambda)} \partial \log(p(x_i|\lambda)/\partial \lambda $ only differ by a  multiplicative constant. In the context of the quantum state $\rho_\lambda$ dependent on the parameter $\lambda$, each measurement $\Pi = \lbrace \Pi_i\rbrace$ performed on the system will lead to the situation above with corresponding probabilities $p_i = \text{Tr}[\Pi_i \rho]$ and CFI $F^C$ depending on the set of probabilities, and thus, ultimately on the measurement $\Pi$. QFI $F^Q$ is simply defined as the supremal CFI $F^C$ over all such measurement bases
\begin{equation}
F^{Q}(\lambda) = \sup_{\Pi} F^C(\lambda)
\end{equation}
It is immediately apparent that finding this optimal measurement allows one to put tighter lower bounds to the variance of the estimator. This is generally the main challenge for quantum metrological problems. A method of obtaining the measurement is via the so-called Symmetric Logarithmic Derivative (SLD) operator $L$ defined as the solution to the matrix equation $\partial_\lambda \rho = \left(\rho L + L \rho\right)/2$.  It can be proved that $F^{Q} = \text{Tr} [\rho L^2]$, thus bypassing the brute-force optimization of the measurement basis. Moreover, an optimal measurement basis can now be shown as just the eigenbasis of the SLD operator $L$. The eigenvectors of the SLD operator therefore always provide an optimal measurement setup that achieves the ultimate precision bound, although it may not be the only basis to achieve that.  

\subsection{QFI of edge states with unconstrained $z$}

In this work, we consider SPT systems, which possess topological features as long as the system Hamiltonian preserves certain symmetries, and these symmetries are said to preserve the topological phase. SPT systems are short-range entangled and do not require presence of interactions. Over the last two decades, many free-fermionic SPT systems have been fabricated on solid state platform, and verified using experimental techniques such as conductance measurements and angle-resolved photo-emission spectroscopy (ARPES). Other experimental platforms including ultracold atoms, trapped ions, and photonics have also achieved great success in simulating SPT phases. In addition to being experimentally accessible, free-fermionic SPT phases are also theoretically tractable due to absence of interactions. In fact, an exhaustive classification of free-fermionic and -bosonic SPT systems in all spatial dimensions and satisfying various symmetry combinations is well understood.

One of the most important feature of free-fermionic SPT systems is the presence of protected energy eigenstates localized on their edges/surfaces. These states play a dominant role in bestowing exotic topological properties, such as the quantized transverse conductance in integer quantum Hall states. Such localized energy states are theoretically investigated using tight-binding models, such as the SSH Hamiltonian and the Chern insulator Hamiltonian considered in this paper. These localized states persist even in the presence of weak disorder satisfying the symmetries protecting the topological phase, and this phenomenon is dubbed as ``bulk-boundary correspondence''. In our work, we propose a sensing procedure that utilizes the localization properties of these edge states. In the topological phase and away from the phase boundary, the edge states are exponentially localized near the boundary. At the topological phase transition, the localization length of the edge states diverges, thus transforming the edge states into extended energy states. We show in the present paper that this transition in the localization length can be used for Heisenberg-limited sensing.

In the main text, the edge state wavefunction is expressed in Eq.~\eqref{generic_edge}, in terms of localization parameter $z$. We have shown the scaling behaviour for the case where $\arg(z)$ is independent of the parameter to be estimated, $\lambda$. Now we derive the scaling of QFI for general case where this constraint is removed. For $z = re^{i\theta}$, we can express $\ket{\partial_\lambda \phi_z} = (\partial_\lambda r) \ket{\partial_r\phi_z} + 
(\partial_\lambda \theta) \ket{\partial_\theta \phi_z}$.
Then the QFI of $\ket{\phi_z}$ can be expressed as
\begin{equation}
F_{\ket{\phi_z}}(\lambda) = \begin{bmatrix}(\partial_\lambda r) & (\partial_\lambda \theta)\end{bmatrix}
\begin{bmatrix}
F_{\ket{\phi_z}}(r,r) & F_{\ket{\phi_z}}(r,\theta) \\
F_{\ket{\phi_z}}(\theta,r) & F_{\ket{\phi_z}}(\theta,\theta)
\end{bmatrix}
\begin{bmatrix}(\partial_\lambda r) \\ (\partial_\lambda \theta)\end{bmatrix} \,,
\end{equation}
where 
\begin{eqnarray}
\label{rtheta}
F_{\ket{\phi_z}}(r,r) &=& 4\Big( \braket{\partial_r\phi_z | \partial_r \phi_z} - \braket{\partial_r\phi_z | \phi_z} \braket{\phi_z | \partial_r\phi_z} \Big),\nonumber\\
F_{\ket{\phi_z}}(r,\theta) &=& 4\Big( \braket{\partial_r\phi_z | \partial_\theta \phi_z} - \braket{\partial_r\phi_z | \phi_z} \braket{\phi_z | \partial_\theta \phi_z} \Big),\nonumber\\
F_{\ket{\phi_z}}(\theta,r) &=& 4\Big( \braket{\partial_\theta \phi_z | \partial_r \phi_z} - \braket{\partial_\theta \phi_z | \phi_z} \braket{\phi_z | \partial_r\phi_z} \Big),\nonumber\\
F_{\ket{\phi_z}}(\theta,\theta) &=& 4\Big( \braket{\partial_\theta \phi_z | \partial_\theta \phi_z} - \braket{\partial_\theta \phi_z | \phi_z} \braket{\phi_z | \partial_\theta \phi_z} \Big) \,.
\end{eqnarray}
We now use an elegant trick to calculate all quantities on the right-hand side of Eq.~\eqref{rtheta}.  
First define $\ket{\tilde{\phi_z}} = \sum_{j\in [L]}z^j\ket{j}$ so that $\ket{\phi_z} = \ket{\tilde{\phi_z}}/\big\|\ket{\tilde{\phi_z}}\big\|$. 
Some straightforward algebra reveals
\begin{equation}
    F_{\ket{\phi_z}}(r,r) = 4\frac{\braket{\partial_r \tilde{\phi_z} | \partial_r \tilde{\phi_z}}} {\big\|\ket{\tilde{\phi_z}}\big\|^2}
    -4\frac{\braket{\partial_r \tilde{\phi_z} | \tilde{\phi_z}}  \braket{\tilde{\phi_z} | \partial_r\tilde{\phi_z}}} {\big\|\ket{\tilde{\phi_z}}\big\|^4} \,,
\end{equation}
and similar identities hold for $F_{\ket{\phi_z}}(r,\theta), F_{\ket{\phi_z}}(\theta,r)$, and $F_{\ket{\phi_z}}(\theta,\theta)$.
Now observe that
\begin{equation}
    \ket{\partial_r {\tilde{\phi_z}}} = \sum_{j \in [L]}jr^{j-1}e^{i\theta j}\ket{j} = \frac{-i}{r}\sum_{j \in [L]}ijr^je^{i\theta j}\ket{j}
     = \frac{-i}{r}\ket{\partial_{\theta} {\tilde{\phi_z}}} \,.
\end{equation}
Therefore, we get
\begin{align}
\label{Fzcomplex}
F_{\ket{\phi_z}}(\lambda) &= F_{\ket{\phi_z}}(r,r)\begin{bmatrix}(\partial_\lambda r) & (\partial_\lambda \theta)\end{bmatrix}
\begin{bmatrix}
1 & -i/r \\
i/r & 1/r^2
\end{bmatrix}
\begin{bmatrix}(\partial_\lambda r) \\ (\partial_\lambda \theta)\end{bmatrix} \nonumber \\
&= F_{\ket{\phi_z}}(r,r)[(\partial_\lambda r)^2+(\partial_\lambda \theta)^2].
\end{align}
Note that $F_{\ket{\phi_z}}(r,r) = F_{\ket{\phi_z}}(r)$, and the latter was calculated in the main text to yield
\begin{equation}
    F_{\ket{\phi_z}}(r) = \frac{4[1+r^{4L}-r^{2L-2}(2r^2+L^2(1-r^2)^2)]}{(1-r^2)^2(1-r^{2L})^2} \,.
\end{equation}
Away from TPT ($\lambda \ne \lambda_{\rm c}$), we have $r < 1$ and therefore $F_{\ket{\phi_z}}(r) \in \Theta(1)$
so that $F_{\ket{\phi_z}}(\lambda) \in \Theta(1)$ as well.
As TPT is approached, ($\lambda \to \lambda_{\rm c}$), we have $r \to 1$. In this limit, 
\begin{equation}
    \lim_{r \to 1}F_{\ket{\phi_z}}(r) = \frac{L^2-1}{3} \,.
\end{equation}
Finally substituting this value in Eq.~\eqref{Fzcomplex}, we obtain
\begin{equation}
 \lim_{\lambda \to \lambda_{\rm c}} F_{\ket{\phi_z}}(\lambda) = \frac{(L^2-1)[(\partial_\lambda r)^2+(\partial_\lambda \theta)^2]}{3}
\implies  F_{\ket{\phi_z}}(\lambda_{\rm c}) \in \Theta(L^2) \,.
\end{equation}
Therefore, it is proven that even for the most general dependence of $z$ on $\lambda$, QFI of the edge state at TPT shows $\Theta(L^2)$ scaling.

\subsection{Optimal measurement for the general edge state}

In this section, we show that the projective measurement described by projectors $\{\ket{j}\bra{j}\otimes\mathds{1},\ j \in [L]\}$
is optimal up to a length-independent prefactor for the general case where $\arg(z)$ depends on $\lambda$.

We first prove that, if a quantum state $\ket{\psi(\lambda)}$ is expressed in a basis $\ket{e_j}$ has all real coefficients,
then QFI $F^Q_{\ket{\psi}}(\lambda) = F^C_{\ket{\psi}}(\lambda)$, with $F^C_{\ket{\psi}}(\lambda)$ denoting 
the CFI with of the probability distribution $\{p_j = |\braket{\psi|e_j}|^2\}$. To prove this statement,
we express $\ket{\psi}$ as $\ket{\psi} = \sum_{j}\sqrt{p_j}\ket{e_j}$, so that 
$\ket{\partial_\lambda \psi} = \sum_j (\partial_\lambda p_j/2\sqrt{p_j})\ket{j}$. 
Note that $\braket{\psi|\partial_\lambda\psi}=0$,
so that $F^Q_{\ket{\psi}}(\lambda) = 4\braket{\partial_\lambda \psi|\partial_\lambda \psi}$. Now
\begin{align}
    F^Q_{\ket{\psi}}(\lambda) =  4\braket{\partial_\lambda \psi|\partial_\lambda \psi}
    = 4\sum_j \frac{(\partial_\lambda p_j)^2}{4p_j} 
    = \sum_j p_j\left(\frac{\partial_\lambda p_j}{p_j}\right)^2
    = \sum_j p_j(\partial_\lambda \ln p_j)^2 
    = F^C_{\ket{\psi}}(\lambda).
\end{align}
Next, observe that this result holds also under a milder assumption on $\ket{\psi}$, namely that
$\braket{\psi|e_j}$ is complex but $\arg(\braket{\psi|e_j})$  is independent of $\lambda$. Using this result for $\ket{\psi} = \ket{\phi_z}$
and $\{\ket{e_j} = \ket{j}\}$, we get QFI $F_{\ket{\phi_z}}(r) = F^C_{\ket{\phi_z}}(r)$.
Then by Eq.\,\eqref{Fzcomplex}, we get for the general case, where $\arg(z)$ also depends on $\lambda$,
\begin{equation}
    F_{\ket{\phi_z}}(\lambda) = F^C_{\ket{\phi_z}}(\lambda)\left(1+\frac{(\partial_\lambda \theta)^2}{r^2(\partial_\lambda r)^2}\right).
\end{equation}
Finally we have QFI for the edge state $F_{\ket{\psi_{\rm edge}}}(\lambda) = F_{\ket{\phi_z}}(\lambda)+F_{\ket{u}}(\lambda)$ and 
$F^C_{\ket{\psi_{\rm edge}}}(\lambda) = F^C_{\ket{\phi_z}}(\lambda)$, which leads to
\begin{equation}
    F_{\ket{\psi_{\rm edge}}}(\lambda) = F^C_{\ket{\psi_{\rm edge}}}(\lambda)\left(1+\frac{(\partial_\lambda \theta)^2}{r^2(\partial_\lambda r)^2}\right) + F_{\ket{u}}(\lambda).
\end{equation}
Observe that $\left(1+\frac{(\partial_\lambda \theta)^2}{r^2(\partial_\lambda r)^2}\right)$ and $F_{\ket{u}}(\lambda)$
do not depend on $L$. 
Hence, for every value of the parameter $\lambda$, the position measurement is optimal up to a constant prefactor 
and an additive constant independent of $L$.

\subsection{QFI of many-body Slater determinant states}

Starting with the expression for the Slter dterminant state $\ket{\Psi}$ in the main text we now calculate the QFI using the standard formula
\begin{equation}
\label{QFI}
    F_{\ket{\Psi}} = 4\left(\braket{\partial_\lambda \Psi|\partial_\lambda \Psi} - |\braket{\partial_\lambda \Psi|\Psi}|^2 \right) \,.
\end{equation}
To calculate the first term, we proceed as
\begin{align}
    \braket{\partial_\lambda \Psi|\partial_\lambda \Psi} &= \frac{1}{N!}\sum_{\sigma,\tau \in {S_N}}\text{sgn}(\sigma\tau)
    \sum_{l,l'=1}^{N} \Big( \bra{\psi_{\tau_1}}\dots\bra{\partial_\lambda \psi_{\tau_l}}\dots\bra{\psi_{\tau_N}} \Big) \enspace
    \Big( \ket{\psi_{\sigma_1}}\dots\bra{\partial_\lambda \psi_{\sigma_{l'}}}\dots\bra{\psi_{\sigma_N}} \Big) \,.
\end{align}
Note that if $l=l'$, then the non-zero terms correspond to $\sigma = \tau$. On the other hand, if $l\ne l'$,
then non-zero terms correspond to $\sigma = \tau$ and $\sigma = \tau (ll')$, where $(ll')$ is the permutation that exchanges the indices $l$ and $l'$. In the latter case, we have $\text{sgn}(\sigma\tau)=-1$. These observations lead to
\begin{align}
\label{firstterm}
    \braket{\partial_\lambda \Psi|\partial_\lambda \Psi} &= \frac{1}{N!} \sum_{\sigma \in {S_N}} \left( \sum_{l=1}^{N} \braket{\partial_\lambda \psi_{\sigma_l}|\partial_\lambda \psi_{\sigma_l}} 
    + \sum_{l\ne l'} \braket{\partial_\lambda \psi_{\sigma_l}|\psi_{\sigma_l}}\braket{\psi_{\sigma_{l'}}|\partial_\lambda \psi_{\sigma_{l'}}} 
    - \sum_{l\ne l'} \braket{\partial_\lambda \psi_{\sigma_l}|\psi_{\sigma_{l'}}}\braket{\psi_{\sigma_{l'}}|\partial_\lambda \psi_{\sigma_{l}}} \right) \nonumber\\
    &=\sum_{l=1}^{N}
    \braket{\partial_\lambda \psi_{l}|\partial_\lambda \psi_{l}} -
    \sum_{l\ne l'}\braket{\partial_\lambda \psi_{l}|\psi_{l}}\braket{\partial_\lambda \psi_{l'}|\psi_{l'}} - 
    \sum_{l\ne l'}\braket{\partial_\lambda \psi_{l}|\psi_{l'}}\braket{\psi_{l'}|\partial_\lambda \psi_{l}} \,.
\end{align}
We similarly calculate the second term on the right-hand side of Eq.~\eqref{QFI}, which yields
\begin{align}
\label{secondterm}
    \braket{\partial_\lambda \Psi| \Psi}^2 = \left(\sum_{l=1}^{N}
    \braket{\partial_\lambda \psi_{l}|\psi_{l}}\right)^2
    = \sum_{l\ne l'}\braket{\partial_\lambda \psi_{l}|\psi_{l}}\braket{\partial_\lambda \psi_{l'}|\psi_{l'}} + 
    \sum_{l}\braket{\partial_\lambda \psi_{l}|\psi_{l}}^2 \,.
\end{align}
Substituting Eqs.~\eqref{firstterm} and \eqref{secondterm} in Eq.~\eqref{QFI} yields
\begin{align}
    F_{\ket{\Psi}} &= 4\left(\sum_{l=1}^{N}
    \braket{\partial_\lambda \psi_{l}|\partial_\lambda \psi_{l}} - 
    \sum_{l\ne l'}\braket{\partial_\lambda \psi_{l}|\psi_{{l'}}}\braket{\psi_{l'}|\partial_\lambda \psi_{l}}+
    \sum_{l}\braket{\partial_\lambda \psi_{l}|\psi_{l}}^2\right) \nonumber \\
    &=4\sum_{l=1}^{N}\braket{\partial_\lambda \psi_{l}|\mathds{1} - P|\partial_\lambda \psi_l}  \,,    
\end{align}
where $P$ is the projector on the occupied states, i.e.~$P = \sum_{l=1}^{N}\ket{\psi_{l}}\bra{\psi_{l}}$. Now observe that $(\mathds{1}-P)\ket{\psi_l} = 0$ for all $l$, hence we can further express
\begin{align}
\label{manybodyqfi}
    F_{\ket{\Psi}} &= 4\text{Tr} \left[(\mathds{1}-P) \left(\sum_{l}\ket{\partial_\lambda \psi_{l}}\bra{\partial_\lambda \psi_{l}}\right) \right] \nonumber \\
    &= 2\text{Tr} \left[(\mathds{1}-P)\partial^2P \right]\,,
\end{align}
which shows that the many-body QFI can be fully expressed only in terms of the projection operator.

\subsection{Many-body QFI scaling at TPT (PBC)}

Here we consider the Bloch Hamiltonian from the main text that model the band-inversion in 1D systems, namely, $ H_k = \alpha k \sigma_x + (\lambda-\lambda_{\rm c})\sigma_z \ $ near the Dirac point $k=0$. We calculate here the QFI for the two lowest $k$-states which are $k=0$ and $k=2\pi/L$. At Dirac point, we have $H_{k=0}{=}(\lambda-\lambda_{\rm c})\sigma_z$. For $\lambda \ge \lambda_{\rm c}$, we have, $\ket{u_{k=0}} = [0 \enspace 1]^T$, and $\ket{\partial_\lambda u_{k=0}} = 0$, resulting in, $F_{\ket{u_{k=0}}} = 0$. We next look at the QFI of $\ket{u_1} = \ket{u_{k=2\pi/L}}$ corresponding to $k=2\pi/L$ which is the closest point to the Dirac point in the Brillouin zone. Here $H_k = \alpha (2\pi/L)\sigma_x + (\lambda-\lambda_{\rm c})\sigma_z$, so that
\begin{equation}
    \ket{u_1} = \begin{bmatrix} \cos(\gamma/2) \\ \sin(\gamma/2)\end{bmatrix}, \enspace \text{and} \enspace 
    \ket{\partial_\lambda u_1} = \frac{\partial_\lambda \gamma}{2}\begin{bmatrix} -\sin(\gamma/2) \\ \cos(\gamma/2)\end{bmatrix} \,,
\end{equation}
with $\tan \gamma = \alpha/(\lambda-\lambda_{\rm c})$. We now obtain $F_{\ket{u_1}} = (\partial_\lambda \gamma)^2$.
By differentiating the expression for $\tan \gamma$, we get
\begin{equation}
    \partial_\lambda \gamma = \frac{-\alpha/L}{(\lambda - \lambda_{\rm c})^2 + \alpha^2/L^2} \implies 
    \lim_{\lambda \to \lambda_{\rm c}} \partial_\lambda \gamma = -\frac{L}{\alpha}.
\end{equation}
Therefore, at TPT, $F_{\ket{u_1}} = L^2/\alpha^2 \in \Theta(L^2)$.

\subsection{Many-body QFI scaling at TPT for SSH Hamiltonian (PBC)}

The SSH Hamiltonian in Eq.~\eqref{SSHham} in the main text can rewritten in the momentum space with the following form of the Bloch Hamiltonian corresponding to momentum $k = 2 \pi \kappa /L$
\begin{equation}
    H_k^{\rm SSH} = - \begin{bmatrix}
    0 & J_1 + J_2 e^{-i 2\pi \kappa/L} \\
    J_1 + J_2 e^{i 2\pi \kappa/L} & 0
    \end{bmatrix} \,.
\end{equation}
For the filled lower band, we have
\begin{equation}
    \ket{u_k} = \frac{1}{\sqrt{2}}\begin{bmatrix} 1 \\ e^{i\phi_k}\end{bmatrix},\quad
    e^{2i\phi_k} =  \frac{\lambda + e^{i2\pi \kappa/L}}{\lambda + e^{-i2\pi \kappa/L}},
\end{equation}
and
\begin{equation}
    \ket{\partial_\lambda u_k} = \frac{\partial_\lambda \phi_k}{\sqrt{2}}\begin{bmatrix} 0 \\ ie^{i\phi_k}\end{bmatrix} \,.
\end{equation}
We obtain 
\begin{equation}
    F_{\ket{u_k}} = 2(\partial_\lambda \phi_k)^2 +4(i\partial_\lambda \phi_k/2)^2 =  (\partial_\lambda \phi_k)^2 \,.
\end{equation}
At the Dirac point, which is $\kappa=L/2$, we obtain $\phi_k = 0$ independent of $\lambda$, and therefore $F_{\ket{u_{N/2}}}=0$.
To proceed further, we can differentiate the expression for $e^{2i\phi_k}$ for $\kappa \ne L/2$, which yields
\begin{align}
    2ie^{2i\phi_k}\partial_\lambda \phi_k = \frac{1}{\lambda + e^{-i2\pi \kappa/L}} - \frac{\lambda + e^{i2\pi \kappa/L}}{(\lambda + e^{-i2\pi \kappa/L})^2} 
                                = - \frac{2i\sin(2\pi \kappa/L)}{(1+\lambda e^{-i2\pi \kappa/L})^2} \,.
\end{align}
Therefore,
\begin{equation}
    \partial_\lambda \phi_k = - \frac{\sin(2\pi \kappa/L)}{1+\lambda^2+2\lambda\cos(2\pi \kappa/L)} \,.
\end{equation}
At $\lambda = 1$, we can further simplify this expression to
\begin{align}
    \partial_\lambda \phi_k &= - \frac{\sin(2\pi \kappa/L)}{2\left(1+\cos(2\pi \kappa/L)\right)} = - \frac{\tan(\pi \kappa/L)}{2} \,,
\end{align}
and therefore $F_{\ket{u_k}} = \tan^2(\pi \kappa/L) / 4$. Finally, the QFI of the many-body ground state is obtained
by summing over QFI of all $\ket{u_k}$ except at the Dirac point, which is $\kappa = L/2$. We therefore get
\begin{align}    
    F_{\rm PBC}^{\rm SSH} (\lambda_{\rm c}) = \sum_{\kappa=1}^{L-1} \frac{\cot^2(\pi \kappa/L)}{4} = \frac{L^2-3L+2}{12} \,,
\end{align}
where the last equation is obtained using Mathematica. We therefore have proven $\Theta(N^2)$ dependence 
for the QFI of $\ket{\Psi^{\rm PBC}_{\rm GS}}$ for the SSH Hamiltonian.

\subsection{Many-body QFI scaling away from TPT for SSH (PBC)}

We have already shown that QFI scales at most linearly away from TPT for the SSH ground state. In the \emph{continuum} limit, i.e., $L \rightarrow \infty$, we may prove this rigorously in the following way. QFI $F_{\rm PBC}^{\text{SSH}}(\lambda)$ is given by $(\partial_\lambda \phi_k)^2$ summed over all modes $\kappa \in [1,L]$, i.e.,
\begin{equation}
    F_{\rm PBC}^{\text{SSH}}(\lambda) = \sum_{\kappa=1}^{L} \left(\frac{\sin(2\pi \kappa/L)}{1+\lambda^2+2\lambda\cos(2\pi \kappa/L)}\right)^2 \,.
\end{equation}
In the continuum limit, this sum may be approximated by an integral if the function is Riemann-Integrable. That is, 
\begin{equation}
    F_{\rm PBC}^{\text{SSH}}(\lambda) = \frac{L}{ \pi}\int_{0}^{\frac{\pi}{2} - \epsilon} \left(\frac{\sin x}{1+\lambda^2+2\lambda\cos x}\right)^2 dx + \frac{L}{2 \pi}\int_{\frac{\pi}{2} + \epsilon}^{2 \pi} \left(\frac{2\sin x}{1+\lambda^2+2\lambda\cos x}\right)^2 dx \,.
\end{equation}
Note that we have divided the domain of integration into two discontinuous domains to reflect the physics at the Dirac Point. It is easy to see that if this integral exists, then QFI scales linearly with  $L$. This is the case away from TPT, where QFI takes the following form \begin{equation}
    \lim_{L \to \infty} \frac{F_{\rm PBC}^{\text{SSH}}(\lambda)}{L} = \begin{cases}
        \frac{1}{2(1-\lambda^2)} &\quad\text{if } \lambda < 1 \,,\\ 
        \frac{1}{2(\lambda^4-\lambda^2)} &\quad\text{if } \lambda > 1 \,,\\ 
     \end{cases} 
\end{equation}
At TPT, the functions are no longer piecewise Riemann Integrable and linear scaling no longer applies. This is consistent with the quadratic scaling derived above.

\subsection{Many-body QFI scaling at TPT for Chern insulator (PBC)}

For the Chern insulator Hamiltonian in Eq.~\eqref{Chham} in the main text, the lower-band eigenvectors are given by,
\begin{align}
    \ket{\psi_{\boldsymbol{k}}} = \ket{\boldsymbol{k}} \ket{u_{\boldsymbol{k}}} \,,
\end{align}
with 
\begin{align}
    \ket{u_{\boldsymbol{k}}} = \sqrt{\frac{B_z + E_{\boldsymbol{k}}}{2E_{\boldsymbol{k}}}}\begin{bmatrix} \frac{B_z - E_{\boldsymbol{k}}}{B_x + i B_y} \\ 1 \end{bmatrix} \,,
\end{align}
where $E_{\boldsymbol{k}} = \sqrt{B_x^2 + B_y^2 + B_z^2}$ is the magnitude of eigenenergy. After a bit of algebra one finds
\begin{align}
    \ket{\partial_\lambda u_{\boldsymbol{k}}} = \sqrt{\frac{B_z + E_{\boldsymbol{k}}}{2E_{\boldsymbol{k}}}} \frac{ E_{\boldsymbol{k}} - B_z}{2E^2_{\boldsymbol{k}}}\begin{bmatrix} \frac{B_z - E_{\boldsymbol{k}}}{B_x + i B_y} \\ 1 \end{bmatrix} \,, 
\end{align}
which is orthogonal to $\ket{u_{\boldsymbol{k}}}$. Therefore we get
\begin{align}
    F_{\rm PBC}^{\rm Ch} (\lambda)  = \displaystyle\sum_{\boldsymbol{k}} \braket{\partial_\lambda u_{\boldsymbol{k}} | \partial_\lambda u_{\boldsymbol{k}}}
               = \displaystyle\sum_{\boldsymbol{k}} \frac{B_x^2 + B_y^2}{4 E^4_{\boldsymbol{k}}} \,.
\end{align}
We use this expression in the main text to extract the scaling of QFI at TPT and away from it.

\end{document}